\documentclass[%
 aip,
 jmp,%
 amsmath,amssymb,
reprint,%
]{revtex4-1}

\usepackage{graphicx}
\usepackage{dcolumn}
\usepackage{bm}

\begin{document}

\preprint{AIP/123-QED}

\title[Angular momentum and torque described with the complex octonion]{Angular momentum and torque described with the complex octonion}

\author{Zi-Hua Weng}
 \email{xmuwzh@xmu.edu.cn.}
\affiliation{
School of Physics and Mechanical \& Electrical Engineering, Xiamen University, Xiamen 361005, China
}%


\date{\today}

\begin{abstract}
The paper aims to adopt the complex octonion to formulate the angular momentum, torque, and force etc in the electromagnetic and gravitational fields. Applying the octonionic representation enables one single definition of angular momentum (or torque, force) to combine some physics contents, which were considered to be independent of each other in the past. J. C. Maxwell used simultaneously two methods, the vector terminology and quaternion analysis, to depict the electromagnetic theory. It motivates the paper to introduce the quaternion space into the field theory, describing the physical feature of electromagnetic and gravitational fields. The spaces of electromagnetic field and of gravitational field can be chosen as the quaternion spaces, while the coordinate component of quaternion space is able to be the complex number. The quaternion space of electromagnetic field is independent of that of gravitational field. These two quaternion spaces may compose one octonion space. Contrarily, one octonion space can be separated into two subspaces, the quaternion space and $S$-quaternion space. In the quaternion space, it is able to infer the field potential, field strength, field source, angular momentum, torque, and force etc in the gravitational field. In the $S$-quaternion space, it is capable of deducing the field potential, field strength, field source, current continuity equation, and electric (or magnetic) dipolar moment etc in the electromagnetic field. The results reveal that the quaternion space is appropriate to describe the gravitational features, including the torque, force, and mass continuity equation etc. The $S$-quaternion space is proper to depict the electromagnetic features, including the dipolar moment and current continuity equation etc. In case the field strength is weak enough, the force and the continuity equation etc can be respectively reduced to that in the classical field theory.
\end{abstract}

\pacs{03.50.-z; 04.50.Kd; 02.10.De; 45.20.da}
\keywords{angular momentum; torque; quaternion; octonion; gravitational field; electromagnetic field}
\maketitle


\section{\label{sec:level1}Introduction}

In 1873, J. C. Maxwell described the physical feature of electromagnetic field with the vector analysis as well as the quaternion analysis. W. R. Hamilton invented the quaternion in 1843. The octonion, as the ordered couple of quaternions, was invented by J. T. Graves and A. Cayley independently and successively. Later the scientists and engineers separated the quaternion into the scalar part and vector part, to facilitate its application in the engineering. In the works about the electromagnetic theory, J. C. Maxwell mingled naturally the quaternion analysis and vector terminology to depict the electromagnetic feature. Similarly the scholars begin to study the physics feature of gravitational field with the algebra of quaternions.

In recent years applying the quaternion to study the electromagnetic feature has been becoming one significant research orientation, and it continues the development trend of gradual deepening and expanding. Some scholars have been studying the electromagnetic and gravitational theories with the quaternion, trying to promote the further progress of these two field theories. H. T. Anastassiu etc \cite{anastassiu} applied the quaternion to describe the electromagnetic feature. S. M. Grusky etc \cite{grusky} adopted the quaternion to research the time-dependent electromagnetic field of the chiral medium in the classical field theory. K. Morita \cite{morita} developed the study of the quaternion field theory. S. Demir \cite{demir1} and M. Tanisli \cite{tanisli} etc applied the bi-quaternion and hyperbolic quaternion to formulate directly the field equations in the classical field theory. Similarly J. G. Winans \cite{winans} described the physics quantities with the quaternion. J. Edmonds \cite{edmonds} utilized the quaternion to depict the wave equation and gravitational theory in the curved space-time. F. A. Doria \cite{doria} adopted the quaternion to research the gravitational theory. A. S. Rawat \cite{rawat} etc discussed the gravitational field equation with the quaternion treatment. V. Majernik \cite{majernik} studied the extended Maxwell-like gravitational field equations with the quaternion. Moreover several of the scientists apply the octonion analysis to study the electromagnetic theory. M. Gogberashvili \cite{gogberashvili} used the octonion to discuss the electromagnetic theory. V. L. Mironov etc \cite{mironov} described the electromagnetic equations and related features with the algebra of octonions. O. P. S. Negi etc \cite{negi2} depicted the Maxwell's equations by means of the octonion. S. Demir \cite{demir2} expressed the gravitational field equations with the octonion. O. P. S. Negi \cite{negi3}, P. S. Bisht \cite{bisht}, B. S. Rajput \cite{rajput}, H. Dehnen \cite{dehnen}, and S. Dangwal \cite{dangwal} etc adopted the quaternion and octonion to explore the dyon in the gravitational and electromagnetic fields. The paper focuses on the application of the complex quaternion to study the angular momentum, torque, and force etc in the electromagnetic and gravitational fields.

In the existing studies of field theory described with the quaternion up to now, the most of researches regard the quaternion as one simplex substitution for the complex number or the vector in the theoretical applications. Obviously the existing quaternion studies so far have not achieved the expected outcome. This is a far cry from the expectation that an entirely new method can bring some new conclusions. The application of quaternion should enlarge the range of definition of some physics concepts, and bring in new perspectives and inferences.

The ordered couple of quaternions compose the octonion (Table I). On the contrary, the octonion is able to be separated into two parts, the quaternion and the $S$-quaternion (short for the second quaternion), and their coordinates can be chosen as the complex numbers. In the paper, the quaternion space and the $S$-quaternion space can be introduced into the field theory, in order to describe the physical feature of gravitational and electromagnetic fields. One will be able to conclude that the quaternion space is suitable to describe the gravitational features, while the $S$-quaternion space is proper to depict the electromagnetic features. This method can deduce the most of conclusions of the electromagnetic and gravitational fields described with the vector.

In the electromagnetic theory described with the complex quaternion, it is able to deduce directly the Maxwell's equations in the classical electromagnetic theory \cite{honig}, without the help of the current continuity equation. In this approach, substituting the $S$-quaternion for the quaternion, one can obtain the same conclusions still. On the basis of this approach, the paper is able to deduce directly the force and the current continuity equation etc further, and turn these physics contents into the indispensable parts of the electromagnetic theory described with the $S$-quaternion. Meanwhile the paper can apply the quaternion to research the gravitational theory, deducing directly the force, torque, energy, and the mass continuity equation etc.

Between the field theory described by the quaternion or $S$-quaternion \cite{weng} with that by the vector terminology, there are quite a number of inferences in common. However this approach believes that splitting artificially the quaternion into two components, the vector part and the scalar part, must cause a few inference differences between two approaches, the quaternion analysis and the vector terminology. The focus is being placed on the depiction discrepancy of electromagnetic features described by the quaternions with that by the vector, including the continuity equation, force, torque, and energy etc.

By comparison with the classical field theory, one can find that the paper has some improvements as follows. (1) Applying one single octonion space to simultaneously describe the physics quantities of the electromagnetic and gravitational fields. The electromagnetic and gravitational fields are not considered as two isolated parts any more. The electromagnetic field and gravitational field can be combined together to become one united field in the theoretical description, to depict simultaneously the physics features of two fields. (2) Combining like-terms of the physics quantity. In the existing researches, the force (or torque, energy) of electromagnetic and gravitational fields possesses several different terms. The paper is able to unify different terms of the force of electromagnetic and gravitational fields into the single formula, and then to analyze the related physics features. (3) Deducing the continuity equation. Being similar to the force, the continuity equation is also one direct deduction of the field theory described with the quaternions. The mass continuity equation and the current continuity equation both are vital deductions and inevitable components of the field theory with the quaternions.

\section{\label{sec:level1}Field Equations}

In the above, the quaternion space $\mathbb{H}_g$ is independent of the $S$-quaternion space $\mathbb{H}_e$ , but these two quaternion spaces, $\mathbb{H}_g$ and $\mathbb{H}_e$ , can combine together to become one octonion space $\mathbb{O}$ . The quaternion space $\mathbb{H}_g$ is suitable to describe the feature of gravitational field, while the $S$-quaternion space $\mathbb{H}_e$ is proper to depict the property of electromagnetic field. Further the octonion space $\mathbb{O}$ is able to represent simultaneously the physics features of these two fields.

In the quaternion space $\mathbb{H}_g$ for the gravitational field, the basis vector is $\mathbb{H}_g = ( \emph{\textbf{i}}_0, \emph{\textbf{i}}_1, \emph{\textbf{i}}_2, \emph{\textbf{i}}_3 )$, the radius vector is $\mathbb{R}_g = i r_0 \emph{\textbf{i}}_0 + \Sigma r_k \emph{\textbf{i}}_k$, and $\textbf{r} = \Sigma r_k \emph{\textbf{i}}_k$. The velocity is $\mathbb{V}_g = i v_0 \emph{\textbf{i}}_0 + \Sigma v_k \emph{\textbf{i}}_k$, and $\textbf{v} = \Sigma v_k \emph{\textbf{i}}_k $. The gravitational potential is $\mathbb{A}_g = i a_0 \emph{\textbf{i}}_0 + \Sigma a_k \emph{\textbf{i}}_k$, and $\textbf{a} = \Sigma a_k \emph{\textbf{i}}_k $. The gravitational strength is $\mathbb{F}_g = f_0 \emph{\textbf{i}}_0 + \Sigma f_k \emph{\textbf{i}}_k$, and $\textbf{f} = \Sigma f_k \emph{\textbf{i}}_k$. The gravitational source is $\mathbb{S}_g = i s_0 \emph{\textbf{i}}_0 + \Sigma s_k \emph{\textbf{i}}_k$, and $\textbf{s} = \Sigma s_k \emph{\textbf{i}}_k$.
Herein the symbol $\circ$ denotes the octonion multiplication. $r_j, ~v_j, ~a_j, ~s_j$, and $f_0$ are all real. $f_k$ is the complex number. $i$ is the imaginary unit. $\emph{\textbf{i}}_0 = 1$. $j = 0, 1, 2, 3$. $k = 1, 2, 3$.

In the $S$-quaternion space $\mathbb{H}_e$ for the electromagnetic field, the basis vector is $\mathbb{H}_e = ( \emph{\textbf{I}}_0, \emph{\textbf{I}}_1, \emph{\textbf{I}}_2, \emph{\textbf{I}}_3 )$, the radius vector is $\mathbb{R}_e = i R_0 \emph{\textbf{I}}_0 + \Sigma R_k \emph{\textbf{I}}_k$, with $\textbf{R} = \Sigma R_k \emph{\textbf{I}}_k$, and $\textbf{R}_0 = R_0 \emph{\textbf{I}}_0$. The velocity is $\mathbb{V}_e = i V_0 \emph{\textbf{I}}_0 + \Sigma V_k \emph{\textbf{I}}_k $, with $\textbf{V} = \Sigma V_k \emph{\textbf{I}}_k$, and $\textbf{V}_0 = V_0 \emph{\textbf{I}}_0$. The electromagnetic potential is $\mathbb{A}_e = i A_0 \emph{\textbf{I}}_0 + \Sigma A_k \emph{\textbf{I}}_k $, with $\textbf{A} = \Sigma A_k \emph{\textbf{I}}_k $, and $\textbf{A}_0 = A_0 \emph{\textbf{I}}_0$. The electromagnetic source is $\mathbb{S}_e = i S_0 \emph{\textbf{I}}_0 + \Sigma S_k \emph{\textbf{I}}_k$, with $\textbf{S} = \Sigma S_k \emph{\textbf{I}}_k$, and $\textbf{S}_0 = S_0 \emph{\textbf{I}}_0$. The electromagnetic strength is $\mathbb{F}_e = F_0 \emph{\textbf{I}}_0 + \Sigma F_k \emph{\textbf{I}}_k$, with $\textbf{F}_0 = F_0 \emph{\textbf{I}}_0 $, and $\textbf{F} = \Sigma F_k \emph{\textbf{I}}_k$. Herein $\mathbb{H}_e = \mathbb{H}_g \circ \emph{\textbf{I}}_0$. $R_j, ~V_j, ~A_j, ~S_j$, and $F_0$ are all real. $F_k$ is the complex number.

These two quaternion spaces, $\mathbb{H}_g$ and $\mathbb{H}_e$, may compose one octonion space, $\mathbb{O} = \mathbb{H}_g + \mathbb{H}_e$. In the octonion space $\mathbb{O}$ for the electromagnetic and gravitational fields, the octonion radius vector is $\mathbb{R} = \mathbb{R}_g + k_{eg} \mathbb{R}_e$, the octonion velocity is $\mathbb{V} = \mathbb{V}_g + k_{eg} \mathbb{V}_e$, with $k_{eg}$ being one coefficient. The octonion field potential is $\mathbb{A} = \mathbb{A}_g + k_{eg} \mathbb{A}_e$, the octonion field strength is $\mathbb{F} = \mathbb{F}_g + k_{eg} \mathbb{F}_e$. From here on out, some physics quantities are extended to the octonion functions, according to the characteristics of octonion. Apparently $\mathbb{V}$, $\mathbb{A}$, and their differential coefficients are all octonion functions of $\mathbb{R}$.

The octonion definition of field strength, $\mathbb{F} = \square \circ \mathbb{A} $ , can be rewritten as,
\begin{equation}
\mathbb{F}_g + k_{eg} \mathbb{F}_e = \square \circ ( \mathbb{A}_g + k_{eg} \mathbb{A}_e )  ~,
\end{equation}
where there are $\mathbb{F}_g = \square \circ \mathbb{A}_g$, and $\mathbb{F}_e = \square \circ \mathbb{A}_e$, according to the coefficient $k_{eg}$ and basis vectors, $\mathbb{H}_g$ and $\mathbb{H}_e$. The quaternion operator is $\square = \emph{i} \emph{\textbf{i}}_0 \partial_0 + \Sigma \emph{\textbf{i}}_k \partial_k$. $ \nabla = \Sigma \emph{\textbf{i}}_k \partial_k$, $ \partial_j = \partial / \partial r_j$. $v_0 = \partial r_0 / \partial t$. Comparing with the Special Theory of Relativity states that $v_0$ is equal to the speed of light $c$ . $t$ is the time.

For the sake of convenience the paper adopts the gauge condition (Table II), $- f_0 = \partial_0 a_0 - \nabla \cdot \textbf{a} = 0$. And then the gravitational strength can be written as $\textbf{f} = \emph{i} \textbf{g} / v_0 + \textbf{b}$. One component of gravitational strength is the gravitational acceleration, $\textbf{g} / v_0 = \partial_0 \textbf{a} + \nabla a_0$. The other is $\textbf{b} = \nabla \times \textbf{a}$, which is similar to the magnetic flux density. Similarly the paper adopts the gauge condition, $- \textbf{F}_0 = \partial_0 \textbf{A}_0 - \nabla \cdot \textbf{A} = 0$. Therefore the electromagnetic strength is $\textbf{F} = \emph{i} \textbf{E} / v_0 + \textbf{B}$. Herein the electric field intensity is $\textbf{E} / v_0 = \partial_0 \textbf{A} + \nabla \circ \textbf{A}_0$, and the magnetic flux density is $\textbf{B} = \nabla \times \textbf{A}$.

The octonion field source $\mathbb{S}$ of the electromagnetic and gravitational fields is defined as,
\begin{eqnarray}
\mu \mathbb{S} && = - ( \emph{i} \mathbb{F} / v_0 + \square )^* \circ  \mathbb{F}
\nonumber \\
&& = \mu_g \mathbb{S}_g + k_{eg} \mu_e \mathbb{S}_e - ( \emph{i} \mathbb{F} / v_0 )^* \circ \mathbb{F} ~,
\end{eqnarray}
where $\mu$, $\mu_g$, and $\mu_e$ are coefficients. $\mu_g < 0$, and $\mu_e > 0$. $*$ denotes the conjugation of octonion.
In the case for one single particle, a comparison with the classical field theory reveals that, $\mathbb{S}_g = m \mathbb{V}_g$, and $\mathbb{S}_e = q \mathbb{V}_e$. $m$ is the mass density, while $q$ is the density of electric charge.

According to the coefficient $k_{eg}$ and the basis vectors, $\mathbb{H}_g$ and $\mathbb{H}_e$, the octonion definition of the field source can be separated into two parts,
\begin{eqnarray}
&& \mu_g \mathbb{S}_g = - \square^* \circ \mathbb{F}_g  ~,
\\
&& \mu_e \mathbb{S}_e = - \square^* \circ \mathbb{F}_e  ~.
\end{eqnarray}

Obviously Eq.(3) is the definition of gravitational source in the quaternion space $\mathbb{H}_g$. Expanding of Eq.(3) yields the gravitational field equations. When $\textbf{b} = 0$ and $\textbf{a} = 0$, one of the gravitational field equations will be degenerated into the Newton's law of universal gravitation in the classical gravitational theory. Meanwhile Eq.(4) is the definition of electromagnetic source in the $S$-quaternion space $\mathbb{H}_e$ . Expanding of Eq.(4) deduces the Maxwell's equations in the classical electromagnetic theory (Appendix A).

On the analogy of the definition of complex coordinate system, one can define the coordinate of octonion, which involves the quaternion and $S$-quaternion simultaneously. In the octonion coordinate system, the octonion physics quantity can be defined as $ \{ ( \emph{i} c_0 + \emph{i} d_0 \textbf{\emph{I}}_0 ) \circ \emph{\textbf{i}}_0 + \Sigma ( c_k + d_k \textbf{\emph{I}}_0^* ) \circ \emph{\textbf{i}}_k \} $. It means that there are the quaternion coordinate $c_k$ and the $S$-quaternion coordinate $d_k \emph{\textbf{I}}_0^*$ for the basis vector $\emph{\textbf{i}}_k$, while the quaternion coordinate $c_0$ and the $S$-quaternion coordinate $d_0 \emph{\textbf{I}}_0$ for the basis vector $\emph{\textbf{i}}_0$. Herein $c_j$ and $d_j$ are all real.

\section{\label{sec:level1}Angular momentum}

In the octonion space, it is able to define the linear momentum from the field source, and then define the angular momentum and the torque accordingly. The octonion linear momentum density, $\mathbb{P} = \mathbb{P}_g + k_{eg} \mathbb{P}_e$, is defined from the octonion field source $\mathbb{S}$ , and is written as,
\begin{equation}
\mathbb{P} = \mu \mathbb{S} / \mu_g ~,
\end{equation}
where $\mathbb{P}_g = \{ \mu_g \mathbb{S}_g - ( \emph{i} \mathbb{F} / v_0 )^* \circ \mathbb{F} \} / \mu_g$, and $\mathbb{P}_e = \mu_e \mathbb{S}_e / \mu_g$. $\mathbb{P}_g = i p_0 + \textbf{p}$. $\textbf{p} = \Sigma p_k \emph{\textbf{i}}_k$. $\mathbb{P}_e = i \textbf{P}_0 + \textbf{P}$. $\textbf{P} = \Sigma P_k \emph{\textbf{I}}_k$. $\textbf{P}_0 = P_0 \emph{\textbf{I}}_0$. In general, the contribution of the tiny term, $\{ \emph{i} \mathbb{F}^* \circ \mathbb{F} / (\mu_g v_0 )\} $, in the above could be neglected.

In the octonion space, the compounding radius vector is $\mathbb{R}^+ = \mathbb{R} + k_{rx} \mathbb{X}$, with $k_{rx}$ being the coefficient. The physics quantity $\mathbb{X}$ is the integral function of field potential $\mathbb{A}$, that is, $\mathbb{A} = i \square^\times \circ \mathbb{X}$. From the compounding radius vector $\mathbb{R}^+$ and the octonion linear momentum density $\mathbb{P}$, the octonion angular momentum density, $\mathbb{L} = (\mathbb{R}^+ )^\times \circ \mathbb{P}$, is defined as
\begin{eqnarray}
\mathbb{L} = (\mathbb{R}_g^+ + k_{eg} \mathbb{R}_e^+ )^\times \circ (\mathbb{P}_g + k_{eg} \mathbb{P}_e )~,
\end{eqnarray}
and
\begin{eqnarray}
\mathbb{R}^+ = \emph{i} r_0^+ + \textbf{r}^+ + k_{eg} (\emph{i} \textbf{R}_0^+ + \textbf{R}^+)~,
\end{eqnarray}
where $\mathbb{R}_g^+ = \mathbb{R}_g + k_{rx} \mathbb{X}_g$. $\mathbb{R}_e^+ = \mathbb{R}_e + k_{rx} \mathbb{X}_e$. $r_j^+ = r_j + k_{rx} x_j$. $R_j^+ = R_j + k_{rx} X_j$. $\textbf{r}^+ = \Sigma r_k^+ \emph{\textbf{i}}_k$. $\textbf{R}_0^+ = R_0^+ \emph{\textbf{I}}_0$, $\textbf{R}^+ = \Sigma R_k^+ \emph{\textbf{I}}_k$. $\mathbb{X} = \mathbb{X}_g + k_{eg} \mathbb{X}_e$. $\mathbb{X}_g = \emph{i} x_0 + \Sigma x_k \emph{\textbf{i}}_k$, $\mathbb{X}_e = \emph{i} X_0 \emph{\textbf{I}}_0 + \Sigma X_k \emph{\textbf{I}}_k$. $x_j$ and $X_j$ are all real. $\times$ denotes the conjugation of complex number.

The above can be expanded further into
\begin{eqnarray}
\mathbb{L} = && \{ (\mathbb{R}_g^+)^\times \circ \mathbb{P}_g + k_{eg}^2 (\mathbb{R}_e^+)^\times \circ \mathbb{P}_e \}
\nonumber
\\
&& + k_{eg} \{ (\mathbb{R}_g^+)^\times \circ \mathbb{P}_e + (\mathbb{R}_e^+)^\times \circ \mathbb{P}_g \}  ~,
\end{eqnarray}
where the part, $\mathbb{L}_g = (\mathbb{R}_g^+)^\times \circ \mathbb{P}_g + k_{eg}^2 (\mathbb{R}_e^+)^\times \circ \mathbb{P}_e $, situates on the quaternion space $\mathbb{H}_g$, while the part, $\mathbb{L}_e =  (\mathbb{R}_g^+)^\times \circ \mathbb{P}_e + (\mathbb{R}_e^+)^\times \circ \mathbb{P}_g $, stays on the $S$-quaternion space $\mathbb{H}_e$.

In the quaternion space $\mathbb{H}_g$, the component $\mathbb{L}_g$ can be expressed as (Appendix B),
\begin{eqnarray}
\mathbb{L}_g = && r_0^+ p_0 + \textbf{r}^+ \cdot \textbf{p} + k_{eg}^2 ( \textbf{R}_0^+ \circ \textbf{P}_0 + \textbf{R}^+ \cdot \textbf{P} )
\nonumber
\\
&&
+ \emph{i}  \{  p_0 \textbf{r}^+ - r_0^+ \textbf{p}  +  k_{eg}^2 ( \textbf{R}^+ \circ \textbf{P}_0 - \textbf{R}_0^+ \circ \textbf{P} ) \}
\nonumber \\
&& + \textbf{r}^+ \times \textbf{p} + k_{eg}^2 \textbf{R}^+ \times \textbf{P} ~,
\end{eqnarray}
where $\mathbb{L}_g = L_{10} + \emph{i} \textbf{L}_1^i + \textbf{L}_1$. $\textbf{L}_1 = \textbf{r}^+ \times \textbf{p} + k_{eg}^2 \textbf{R}^+ \times \textbf{P} $, includes the angular momentum density. $L_{10}$ covers the dot product of the radius vector and the linear momentum density. $\textbf{L}_1 = \Sigma L_{1k} \emph{\textbf{i}}_k$, $\textbf{L}_1^i = \Sigma L_{1k}^i \emph{\textbf{i}}_k$. $L_{1j}$ and $L_{1k}^i$ are all real.

In the $S$-quaternion space $\mathbb{H}_e$, the component $\mathbb{L}_e$ can be written as,
\begin{eqnarray}
\mathbb{L}_e = && r_0^+ \textbf{P}_0 + \textbf{r}^+ \cdot \textbf{P} + p_0 \textbf{R}_0^+ + \textbf{R}^+ \cdot \textbf{p}
\nonumber
\\
&&
+ \emph{i} (  - r_0^+ \textbf{P} + \textbf{r}^+ \circ \textbf{P}_0  - \textbf{R}_0^+ \circ \textbf{p} + p_0 \textbf{R}^+  )
\nonumber
\\
&&
+  \textbf{r}^+ \times \textbf{P} + \textbf{R}^+ \times \textbf{p} ~,
\end{eqnarray}
where $\mathbb{L}_e = \textbf{L}_{20} + \emph{i} \textbf{L}_2^i + \textbf{L}_2$. $\textbf{L}_2^i$ covers the electric dipole moment etc. $\textbf{L}_2 = \textbf{r}^+ \times \textbf{P} + \textbf{R}^+ \times \textbf{p}$, includes the magnetic dipole moment etc. $\textbf{L}_{20}$ is similar to $L_{10}$ . $\textbf{L}_{20} = L_{20} \emph{\textbf{I}}_0$, $\textbf{L}_2 = \Sigma L_{2k} \emph{\textbf{I}}_k$, $\textbf{L}_2^i = \Sigma L_{2k}^i \emph{\textbf{I}}_k$. $L_{2j}$ and $L_{2k}^i$ are all real.

It means that the definition of octonion angular momentum is able to contain more physics contents, which were considered to be independent of each other in the past, such as the angular momentum and the dipole moment etc. The angular momentum includes the orbital angular momentum etc, while the dipole moment covers the magnetic dipole moment and the electric dipole moment etc.

\section{\label{sec:level1}Octonion Torque}

Making use of the octonion angular momentum, some existing energy terms can be written into a single definition of energy in the octonion space. From the octonion angular momentum density $\mathbb{L}$, the octonion torque density $\mathbb{W}$ can be defined as follows,
\begin{equation}
\mathbb{W} = - v_0 ( \emph{i} \mathbb{F} / v_0 + \square ) \circ \mathbb{L} ~,
\end{equation}
where $\mathbb{L} = \mathbb{L}_g + k_{eg} \mathbb{L}_e$. $\mathbb{W} = \mathbb{W}_g + k_{eg} \mathbb{W}_e$.

The above can be expanded into (Appendix C),
\begin{eqnarray}
\mathbb{W} = && - ( \emph{i} \mathbb{F}_g \circ \mathbb{L}_g + \emph{i} k_{eg}^2 \mathbb{F}_e \circ \mathbb{L}_e + v_0 \square \circ \mathbb{L}_g )
\nonumber
\\
&&
- k_{eg} ( \emph{i} \mathbb{F}_g \circ \mathbb{L}_e + \emph{i} \mathbb{F}_e \circ \mathbb{L}_g + v_0 \square \circ \mathbb{L}_e )~,
\end{eqnarray}
where the component, $\mathbb{W}_g = - ( \emph{i} \mathbb{F}_g \circ \mathbb{L}_g + \emph{i} k_{eg}^2 \mathbb{F}_e \circ \mathbb{L}_e + v_0 \square \circ \mathbb{L}_g ) $, situates on the quaternion space $\mathbb{H}_g$, while the component, $\mathbb{W}_e = - ( \emph{i} \mathbb{F}_g \circ \mathbb{L}_e + \emph{i} \mathbb{F}_e \circ \mathbb{L}_g + v_0 \square \circ \mathbb{L}_e ) $, stays on the $S$-quaternion space $\mathbb{H}_e$.

\subsection{\label{sec:level2}Component $\mathbb{W}_g$}

In the quaternion space $\mathbb{H}_g$, the component $\mathbb{W}_g$ can be expressed as,
\begin{eqnarray}
\mathbb{W}_g = &&
\{
( \textbf{g} L_{10} / v_0 + \textbf{g} \times \textbf{L}_1 / v_0 + \textbf{b} \times \textbf{L}_1^i )
\nonumber
\\
&&
- v_0 ( - \partial_0 \textbf{L}_1^i + \nabla L_{10} +  \nabla \times \textbf{L}_1 )
\nonumber \\
&& + k_{eg}^2 ( \textbf{E} \circ \textbf{L}_{20} / v_0 + \textbf{E} \times \textbf{L}_2 / v_0 + \textbf{B} \times \textbf{L}_2^i )
\}
\nonumber
\\
&& +  \emph{i} \{ ( \textbf{g} \cdot \textbf{L}_1^i / v_0 - \textbf{b} \cdot \textbf{L}_1 )
- v_0 (  \partial_0 L_{10} +  \nabla \cdot \textbf{L}_1^i)
\nonumber
\\
&&
+ k_{eg}^2 ( \textbf{E} \cdot \textbf{L}_2^i / v_0 - \textbf{B} \cdot \textbf{L}_2 )
\}
\nonumber
\\
&& + \emph{i} \{ ( \textbf{g} \times \textbf{L}_1^i / v_0 - L_{10} \textbf{b} - \textbf{b} \times \textbf{L}_1 )
\nonumber \\
&&
- v_0 (  \partial_0 \textbf{L}_1 +  \nabla \times \textbf{L}_1^i )
\nonumber \\
&& + k_{eg}^2 ( \textbf{E} \times \textbf{L}_2^i / v_0 - \textbf{B} \circ \textbf{L}_{20} - \textbf{B} \times \textbf{L}_2 )
\}
\nonumber
\\
&& + \{ ( \textbf{b} \cdot \textbf{L}_1^i + \textbf{g} \cdot \textbf{L}_1 / v_0 )
- v_0 ( \nabla \cdot \textbf{L}_1 )
\nonumber
\\
&&
+ k_{eg}^2 ( \textbf{B} \cdot \textbf{L}_2^i + \textbf{E} \cdot \textbf{L}_2 / v_0 ) \}~,
\end{eqnarray}
where $\mathbb{W}_g = \emph{i} W_{10}^i + W_{10} + \emph{i} \textbf{W}_1^i + \textbf{W}_1$. $W_{10}^i$ is the energy density. The energy includes the proper energy, kinetic energy, potential energy, work, and the interacting energy between the electric field intensity with the electric dipole moment, and the interacting energy between the magnetic flux density with the magnetic dipole moment. $-\textbf{W}_1^i$ is the torque density, covering the torque density produced by the applied force etc. $\textbf{W}_1$ is the curl of angular momentum density, while $W_{10}$ is the divergence of angular momentum density. $\textbf{W}_1 = \Sigma W_{1k} \emph{\textbf{i}}_k$, $\textbf{W}_1^i = \Sigma W_{1k}^i \emph{\textbf{i}}_k$. $W_{1j}$ and $W_{1j}^i$ are all real.

The expression of energy $W_{10}^i$ is associated with the definition of field potential. The gravitational potential, $\mathbb{A}_g = \emph{i} a_0 + \textbf{a}$, is defined as, $\mathbb{A}_g = \emph{i} \square^\times \circ \mathbb{X}_g$, with $a_0 = \partial_0 x_0 + \nabla \cdot \textbf{x}$, and $\textbf{a} = \partial_0 \textbf{x} - \nabla x_0$. The electromagnetic potential, $\mathbb{A}_e = \emph{i} \textbf{A}_0 + \textbf{A}$, is defined as, $\mathbb{A}_e = \emph{i} \square^\times \circ \mathbb{X}_e$, with $\textbf{A}_0 = \partial_0 \textbf{X}_0 + \nabla \cdot \textbf{X}$, and $\textbf{A} = \partial_0 \textbf{X} - \nabla \circ \textbf{X}_0$. The gauge conditions are chosen as, $\nabla \times \textbf{x} = 0$, and $\nabla \times \textbf{X} = 0$.

The energy density $W_{10}^i$ is written as
\begin{eqnarray}
W_{10}^i = && ( \textbf{g} \cdot \textbf{L}_1^i / v_0 - \textbf{b} \cdot \textbf{L}_1 )
\nonumber
\\
&&
- v_0 ( \partial_0 L_{10} + \nabla \cdot \textbf{L}_1^i)
\nonumber
\\
&&
+ k_{eg}^2 ( \textbf{E} \cdot \textbf{L}_2^i / v_0 - \textbf{B} \cdot \textbf{L}_2 )
\nonumber \\
\approx
&& - \{ v_0 p_0 + v_0 p_0 ( \nabla \cdot \textbf{r} ) \}
\nonumber \\
&&
- \{ v_0 ( \partial_0 \textbf{r}) \cdot \textbf{p} + v_0 \textbf{r} \cdot \partial_0 \textbf{p} \}
\nonumber \\
&&
- \{ ( p_0 a_0 + \textbf{a} \cdot \textbf{p})
\nonumber \\
&&
+ k_{eg}^2 ( \textbf{A}_0 \circ \textbf{P}_0 + \textbf{A} \cdot \textbf{P} ) \}
\nonumber
\\
&&
+  k_{eg}^2 \{ (\textbf{E} / v_0 ) \cdot ( \textbf{r} \circ \textbf{P}_0 )
\nonumber \\
&&
- \textbf{B} \cdot ( \textbf{r} \times \textbf{P} ) \}
\nonumber \\
&&
+ \{ ( \textbf{g} / v_0 ) \cdot ( p_0 \textbf{r} ) - \textbf{b} \cdot \textbf{L}_1 \}
~,
\end{eqnarray}
where $- \{ v_0 p_0 + v_0 p_0 ( \nabla \cdot \textbf{r} ) \} = k_p v_0 p_0$, covers the proper energy $v_0 s_0$ etc. $- \{ v_0 ( \partial_0 \textbf{r}) \cdot \textbf{p} + v_0 \textbf{r} \cdot \partial_0 \textbf{p} \}$ is the sum of the kinetic energy and the work produced by the applied force. $- \{ ( p_0 a_0 + \textbf{a} \cdot \textbf{p}) + k_{eg}^2 ( \textbf{A}_0 \circ \textbf{P}_0 + \textbf{A} \cdot \textbf{P} ) \}$ is the potential energy of gravitational and electromagnetic fields. $\{ k_{eg}^2  (\textbf{E} / v_0 ) \cdot ( \textbf{r} \circ \textbf{P}_0 ) \}$ is the interacting energy between the electric field intensity with the electric dipole moment. $\{ - k_{eg}^2 \textbf{B} \cdot ( \textbf{r} \times \textbf{P} ) \}$ is the interacting energy between the magnetic flux density with the magnetic dipole moment. $( - \textbf{b} \cdot \textbf{L}_1 )$ can be considered as the interacting energy between the gravitational field with the angular momentum.
$k_p = (k - 1)$ is the coefficient, with $k$ being the dimension of vector $\textbf{r}$. For the case of $k = 3$, $( W_{10}^i / 2 )$ is the conventional energy in the classical field theory. And it is stated, $k_{rx} = 1 / v_0$, to compare with the potential energy in the classical gravitational and electromagnetic fields.

The above states that, in case the compounding radius vector $\mathbb{R}^+$ neglects the integral function $\mathbb{X}$, the energy definition in the quaternion space $\mathbb{H}_g$ will be incomplete. Further it will result in the energy definition to exclude the potential energy of gravitational and electromagnetic fields etc.

Similarly the torque density $-\textbf{W}_1^i$ is written as,
\begin{eqnarray}
\textbf{W}_1^i = && ( \textbf{g} \times \textbf{L}_1^i / v_0 - L_{10} \textbf{b} - \textbf{b} \times \textbf{L}_1 )
\nonumber \\
&&
- v_0 ( \partial_0 \textbf{L}_1 + \nabla \times \textbf{L}_1^i )
\nonumber \\
&& + k_{eg}^2 ( \textbf{E} \times \textbf{L}_2^i / v_0 - \textbf{B} \circ \textbf{L}_{20} - \textbf{B} \times \textbf{L}_2 )
\nonumber \\
\approx &&  p_0 \textbf{g} \times \textbf{r} / v_0 - v_0 \textbf{r} \times \partial_0 \textbf{p}
\nonumber \\
&& + k_{eg}^2 \{ \textbf{E} \times ( \textbf{r} \circ \textbf{P}_0 ) / v_0  -  \textbf{B} \times ( \textbf{r} \times \textbf{P} )  \}
\nonumber \\
&&
- ( \textbf{r} \cdot \textbf{p} ) \textbf{b} - \textbf{b} \times ( \textbf{r} \times \textbf{p} ) + \textbf{a} \times \textbf{p} ~,
\end{eqnarray}
where $( - p_0 \textbf{g} \times \textbf{r} / v_0 )$ is the torque term produced by the gravity. And $( - v_0 \textbf{r} \times \partial_0 \textbf{p} )$ is the torque term produced by the inertial force. $\{ - k_{eg}^2 \textbf{E} \times ( \textbf{r} \circ \textbf{P}_0 ) / v_0 \}$ is the torque term caused by the electric field intensity and electric dipole moment. $\{ k_{eg}^2 \textbf{B} \times ( \textbf{r} \times \textbf{P} ) \}$ is the torque term caused by the magnetic flux density and the magnetic dipole moment.

\begin{table}[h]
\caption{The multiplication table of octonion.}
\label{tab:table3}
\centering
\begin{ruledtabular}
\begin{tabular}{ccccccccc}
$ $ & $1$ & $\emph{\textbf{i}}_1$  &
$\emph{\textbf{i}}_2$ & $\emph{\textbf{i}}_3$  &
$\emph{\textbf{I}}_0$ & $\emph{\textbf{I}}_1$
& $\emph{\textbf{I}}_2$  & $\emph{\textbf{I}}_3$  \\
\hline $1$ & $1$ & $\emph{\textbf{i}}_1$  & $\emph{\textbf{i}}_2$ &
$\emph{\textbf{i}}_3$  & $\emph{\textbf{I}}_0$  &
$\emph{\textbf{I}}_1$
& $\emph{\textbf{I}}_2$  & $\emph{\textbf{I}}_3$  \\
$\emph{\textbf{i}}_1$ & $\emph{\textbf{i}}_1$ & $-1$ &
$\emph{\textbf{i}}_3$  & $-\emph{\textbf{i}}_2$ &
$\emph{\textbf{I}}_1$
& $-\emph{\textbf{I}}_0$ & $-\emph{\textbf{I}}_3$ & $\emph{\textbf{I}}_2$  \\
$\emph{\textbf{i}}_2$ & $\emph{\textbf{i}}_2$ &
$-\emph{\textbf{i}}_3$ & $-1$ & $\emph{\textbf{i}}_1$  &
$\emph{\textbf{I}}_2$  & $\emph{\textbf{I}}_3$
& $-\emph{\textbf{I}}_0$ & $-\emph{\textbf{I}}_1$ \\
$\emph{\textbf{i}}_3$ & $\emph{\textbf{i}}_3$ &
$\emph{\textbf{i}}_2$ & $-\emph{\textbf{i}}_1$ & $-1$ &
$\emph{\textbf{I}}_3$  & $-\emph{\textbf{I}}_2$
& $\emph{\textbf{I}}_1$  & $-\emph{\textbf{I}}_0$ \\

$\emph{\textbf{I}}_0$ & $\emph{\textbf{I}}_0$ &
$-\emph{\textbf{I}}_1$ & $-\emph{\textbf{I}}_2$ &
$-\emph{\textbf{I}}_3$ & $-1$ & $\emph{\textbf{i}}_1$
& $\emph{\textbf{i}}_2$  & $\emph{\textbf{i}}_3$  \\
$\emph{\textbf{I}}_1$ & $\emph{\textbf{I}}_1$ &
$\emph{\textbf{I}}_0$ & $-\emph{\textbf{I}}_3$ &
$\emph{\textbf{I}}_2$  & $-\emph{\textbf{i}}_1$
& $-1$ & $-\emph{\textbf{i}}_3$ & $\emph{\textbf{i}}_2$  \\
$\emph{\textbf{I}}_2$ & $\emph{\textbf{I}}_2$ &
$\emph{\textbf{I}}_3$ & $\emph{\textbf{I}}_0$  &
$-\emph{\textbf{I}}_1$ & $-\emph{\textbf{i}}_2$
& $\emph{\textbf{i}}_3$  & $-1$ & $-\emph{\textbf{i}}_1$ \\
$\emph{\textbf{I}}_3$ & $\emph{\textbf{I}}_3$ &
$-\emph{\textbf{I}}_2$ & $\emph{\textbf{I}}_1$  &
$\emph{\textbf{I}}_0$  & $-\emph{\textbf{i}}_3$
& $-\emph{\textbf{i}}_2$ & $\emph{\textbf{i}}_1$  & $-1$ \\
\end{tabular}
\end{ruledtabular}
\end{table}

\subsection{\label{sec:level2}Component $\mathbb{W}_e$}

In the $S$-quaternion space $\mathbb{H}_e$, the component $\mathbb{W}_e$ can be written as,
\begin{eqnarray}
\mathbb{W}_e = \emph{i} \textbf{W}_{20}^i + \textbf{W}_{20} + \emph{i} \textbf{W}_2^i + \textbf{W}_2~,
\end{eqnarray}
where $\textbf{W}_{20}$ covers the divergence of magnetic dipole moment, while $\textbf{W}_2$ includes the curl of magnetic dipole moment and the derivative of electric dipole moment. $\textbf{W}_{20} = W_{20} \emph{\textbf{I}}_0$, $\textbf{W}_{20}^i = W_{20}^i \emph{\textbf{I}}_0$. $\textbf{W}_2 = \Sigma W_{2k} \emph{\textbf{I}}_k$, $\textbf{W}_2^i = \Sigma W_{2k}^i \emph{\textbf{I}}_k$. $W_{2j}$ and $W_{2j}^i$ are all real. That is,
\begin{eqnarray}
\textbf{W}_{20}^i = && ( \textbf{g} \cdot \textbf{L}_2^i / v_0 - \textbf{b} \cdot \textbf{L}_2 )
                    \nonumber \\
                    &&
                    + ( \textbf{E} \cdot \textbf{L}_1^i / v_0 - \textbf{B} \cdot \textbf{L}_1 )
                    \nonumber \\
                    && - v_0 ( \partial_0 \textbf{L}_{20} + \nabla \cdot \textbf{L}_2^i ) ~,
\\
\textbf{W}_{20} = && ( \textbf{b} \cdot \textbf{L}_2^i + \textbf{g} \cdot \textbf{L}_2 / v_0 ) - v_0 ( \nabla \cdot \textbf{L}_2 )
                    \nonumber \\
                    && + ( \textbf{B} \cdot \textbf{L}_1^i + \textbf{E} \cdot \textbf{L}_1 / v_0 ) ~,
\\
\textbf{W}_2^i = && ( \textbf{g} \times \textbf{L}_2^i / v_0 - \textbf{b} \circ \textbf{L}_{20} - \textbf{b} \times \textbf{L}_2 )
                    \nonumber \\
                    && + ( \textbf{E} \times \textbf{L}_1^i / v_0 - L_{10} \textbf{B} - \textbf{B} \times \textbf{L}_1 )
                    \nonumber \\
                    && - v_0 ( \partial_0 \textbf{L}_2 + \nabla \times \textbf{L}_2^i ) ~,
\\
\textbf{W}_2 = && ( \textbf{g} \circ \textbf{L}_{20} / v_0 + \textbf{g} \times \textbf{L}_2 / v_0 + \textbf{b} \times \textbf{L}_2^i )
                   \nonumber \\
                   && + ( L_{10} \textbf{E} / v_0 + \textbf{E} \times \textbf{L}_1 / v_0 + \textbf{B} \times \textbf{L}_1^i )
                   \nonumber \\
                   && - v_0 ( - \partial_0 \textbf{L}_2^i + \nabla \circ \textbf{L}_{20} + \nabla \times \textbf{L}_2 ) ~.
\end{eqnarray}

The above means that it is comparative easy to distinguish the physics quantity in the gravitational field with that in the electromagnetic field, in the definitions for the field potential, field strength, and field source. However in the definitions for the angular momentum, energy, and torque, the physics quantities of those two fields are mixed more and more complicatedly. Those mixed physics quantities have the direct influence on some physics quantities of each field (Table III).

The paper deals with not only the physics quantity in the subspace $\mathbb{H}_g$ but also that in the subspace $\mathbb{H}_e$ . It is noticeable that the physics quantity in the subspace $\mathbb{H}_e$ is able to impact directly the multiplication and related differential coefficient in the subsequent deductions, and then should not be neglected artificially. For example, the terms $\textbf{W}_{20}^i$ and $\textbf{W}_2^i$ etc of the physics quantity $\mathbb{W}_e$ in the subspace $\mathbb{H}_e$ have the direct contribution to the power and force of the physics quantity $\mathbb{N}$ in the following section, and the contribution of those terms is able to be measured directly.

It states that the definition of octonion torque is able to combine some physics contents, which were considered to be independent of each other before, such as the energy, the torque, the divergence of and curl of magnetic dipole moment, and the derivative of electric dipole moment etc. The energy covers the proper energy, kinetic energy, work, and potential energy etc. And the torque includes the torque between the electric field intensity with the electric dipole moment, and the torque between the magnetic flux density with the magnetic dipole moment (or the spin magnetic moment) etc.

\begin{table}[h]
\caption{The multiplication of the operator and octonion physics quantity.}
\label{tab:table2}
\centering
\begin{ruledtabular}
\begin{tabular}{ll}
definition                  &   expression~meaning                                   \\
\hline
$\nabla \cdot \textbf{a}$   &  $-(\partial_1 a_1 + \partial_2 a_2 + \partial_3 a_3)$ \\
$\nabla \times \textbf{a}$  &  $\emph{\textbf{i}}_1 ( \partial_2 a_3
                                 - \partial_3 a_2 ) + \emph{\textbf{i}}_2 ( \partial_3 a_1
                                 - \partial_1 a_3 )$                                 \\
$$                          &  ~~~$+ \emph{\textbf{i}}_3 ( \partial_1 a_2
                                 - \partial_2 a_1 )$                                 \\
$\nabla a_0$                &  $\emph{\textbf{i}}_1 \partial_1 a_0
                                 + \emph{\textbf{i}}_2 \partial_2 a_0
                                 + \emph{\textbf{i}}_3 \partial_3 a_0  $             \\
$\partial_0 \textbf{a}$     &  $\emph{\textbf{i}}_1 \partial_0 a_1
                                 + \emph{\textbf{i}}_2 \partial_0 a_2
                                 + \emph{\textbf{i}}_3 \partial_0 a_3  $             \\

$\nabla \cdot \textbf{A}$   &  $-(\partial_1 A_1 + \partial_2 A_2 + \partial_3 A_3) \emph{\textbf{I}}_0 $  \\
$\nabla \times \textbf{A}$  &  $-\emph{\textbf{I}}_1 ( \partial_2
                                 A_3 - \partial_3 A_2 ) - \emph{\textbf{I}}_2 ( \partial_3 A_1
                                 - \partial_1 A_3 )$                                 \\
$$                          &  ~~~$- \emph{\textbf{I}}_3 ( \partial_1 A_2
                                 - \partial_2 A_1 )$                                 \\
$\nabla \circ \textbf{A}_0$ &  $\emph{\textbf{I}}_1 \partial_1 A_0
                                 + \emph{\textbf{I}}_2 \partial_2 A_0
                                 + \emph{\textbf{I}}_3 \partial_3 A_0  $             \\
$\partial_0 \textbf{A}$     &  $\emph{\textbf{I}}_1 \partial_0 A_1
                                 + \emph{\textbf{I}}_2 \partial_0 A_2
                                 + \emph{\textbf{I}}_3 \partial_0 A_3  $             \\
\end{tabular}
\end{ruledtabular}
\end{table}

\section{\label{sec:level1}Octonion Force}

Expanding the defining range of angular momentum and of torque will result in the expansion of the defining range of force correspondingly. One of the consequences will enable a single definition of force to contain different force terms as much as possible. In other words, some existing force terms can be united into the same one definition in the octonion space.

From the torque density $\mathbb{W}$, the octonion force density $\mathbb{N}$ can be defined as follows,
\begin{equation}
\mathbb{N} = - ( \emph{i} \mathbb{F} / v_0 + \square ) \circ \mathbb{W},
\end{equation}
where $\mathbb{N} = \mathbb{N}_g + k_{eg} \mathbb{N}_e$.

Further the above can be expanded into (Appendix D),
\begin{eqnarray}
\mathbb{N} = && - ( \square \circ \mathbb{W}_g + \emph{i} \mathbb{F}_g \circ \mathbb{W}_g / v_0
\nonumber \\
&&
+ \emph{i} k_{eg}^2 \mathbb{F}_e \circ \mathbb{W}_e / v_0 )
\nonumber \\
&& - k_{eg} ( \square \circ \mathbb{W}_e + \emph{i} \mathbb{F}_g \circ \mathbb{W}_e / v_0
\nonumber \\
&&
+ \emph{i} \mathbb{F}_e \circ \mathbb{W}_g / v_0 ),
\end{eqnarray}
where the component $\mathbb{N}_g = - ( \emph{i} \mathbb{F}_g \circ \mathbb{W}_g / v_0 + \emph{i} k_{eg}^2 \mathbb{F}_e \circ \mathbb{W}_e / v_0 + \square \circ \mathbb{W}_g ) $ situates on the quaternion space $\mathbb{H}_g$, while the component $\mathbb{N}_e = - ( \emph{i} \mathbb{F}_g \circ \mathbb{W}_e / v_0 + \emph{i} \mathbb{F}_e \circ \mathbb{W}_g / v_0 + \square \circ \mathbb{W}_e ) $ stays on the $S$-quaternion space $\mathbb{H}_e$.

\subsection{\label{sec:level2}Component $\mathbb{N}_g$}

In the quaternion space $\mathbb{H}_g$, the component $\mathbb{N}_g$ can be expressed as,
\begin{eqnarray}
\mathbb{N}_g = &&
\emph{i}
\{
( W_{10}^i \textbf{g} / v_0 + \textbf{g} \times \textbf{W}_1^i / v_0
\nonumber \\
&&
- W_{10} \textbf{b} - \textbf{b} \times \textbf{W}_1 ) / v_0
\nonumber \\
&& + k_{eg}^2 ( \textbf{E} \circ \textbf{W}_{20}^i / v_0 + \textbf{E} \times \textbf{W}_2^i / v_0
\nonumber \\
&&
- \textbf{B} \circ \textbf{W}_{20} - \textbf{B} \times \textbf{W}_2 ) / v_0
\nonumber \\
&& - ( \partial_0 \textbf{W}_1 + \nabla W_{10}^i + \nabla \times \textbf{W}_1^i )
\}
\nonumber
\\
&& + \{ ( W_{10} \textbf{g} / v_0 + \textbf{g} \times \textbf{W}_1 / v_0
\nonumber \\
&&
+ W_{10}^i \textbf{b} + \textbf{b} \times \textbf{W}_1^i ) / v_0
\nonumber \\
&& + k_{eg}^2 ( \textbf{E} \circ \textbf{W}_{20} / v_0 + \textbf{E} \times \textbf{W}_2/ v_0
\nonumber \\
&&
+ \textbf{B} \circ \textbf{W}_{20}^i + \textbf{B} \times \textbf{W}_2^i ) / v_0
\nonumber \\
&& + ( \partial_0 \textbf{W}_1^i - \nabla W_{10} - \nabla \times \textbf{W}_1 )
\}
\nonumber
\\
&& + \emph{i} \{ ( \textbf{g} \cdot \textbf{W}_1^i / v_0 - \textbf{b} \cdot \textbf{W}_1 ) / v_0
\nonumber \\
&&
+  k_{eg}^2 ( \textbf{E} \cdot \textbf{W}_2^i / v_0 - \textbf{B} \cdot \textbf{W}_2 ) / v_0
\nonumber \\
&&
- ( \partial_0 W_{10} + \nabla \cdot \textbf{W}_1^i )
\}
\nonumber
\\
&& + \{ ( \textbf{g} \cdot \textbf{W}_1 / v_0 + \textbf{b} \cdot \textbf{W}_1^i ) / v_0
\nonumber \\
&&
+ k_{eg}^2 ( \textbf{E} \cdot \textbf{W}_2/ v_0 + \textbf{B} \cdot \textbf{W}_2^i ) / v_0
\nonumber \\
&&
+ ( \partial_0 W_{10}^i -  \nabla \cdot \textbf{W}_1 )
\}~  ,
\end{eqnarray}
where $\mathbb{N}_g = \emph{i} N_{10}^i + N_{10} + \emph{i} \textbf{N}_1^i + \textbf{N}_1 $ . $N_{10}$ is the power density, which is able to figure out the mass continuity equation. $N_{10}^i$ covers the torque divergence. $\textbf{N}_1^i$ is the force density. $\textbf{N}_1$ is the torque derivative. $\textbf{N}_1 = \Sigma N_{1k} \emph{\textbf{i}}_k$, $\textbf{N}_1^i = \Sigma N_{1k}^i \emph{\textbf{i}}_k$.  $N_{1j}$ and $N_{1j}^i$ are all real.

When $\mathbb{N}_g = 0$, the mass continuity equation can be derived from $N_{10} = 0$. From $\textbf{N}_1 = 0$, one can deduce the velocity curl of the particle. Especially the angular velocity of Larmor precession for one charged particle can be derived from $\textbf{N}_1 = 0$. The force equilibrium equation is able to be deduced from $\textbf{N}_1^i = 0$. Further $\textbf{N}_1^i = 0$ deduces that the gravitational acceleration $\textbf{g}$ corresponds to the linear acceleration. And $\textbf{N}_1 = 0$ infers that the component $\textbf{b}$ of gravitational strength corresponds to the velocity curl, which is the double of the precessional angular velocity when $k = 2$.

In case the field strength is relatively weak, the definition of power density,
\begin{eqnarray}
N_{10} = && ( \textbf{g} \cdot \textbf{W}_1 / v_0 + \textbf{b} \cdot \textbf{W}_1^i ) / v_0
\nonumber \\
&&
+ ( \partial_0 W_{10}^i - \nabla \cdot \textbf{W}_1 )
\nonumber \\
&&
+ k_{eg}^2 ( \textbf{E} \cdot \textbf{W}_2 / v_0 + \textbf{B} \cdot \textbf{W}_2^i ) / v_0
~,
\end{eqnarray}
can be written approximately as,
\begin{eqnarray}
N_{10} / k_p \approx && \partial_0 ( p_0 v_0 ) - \nabla \cdot ( \textbf{p} v_0 )
\nonumber \\
&&
+ L_{10} ( \textbf{b} \cdot \textbf{b} - \textbf{g} \cdot \textbf{g} / v_0^2 ) / ( v_0 k_p )
\nonumber \\
&&
+ k_{eg}^2 L_{10} ( \textbf{B} \cdot \textbf{B} - \textbf{E} \cdot \textbf{E} / v_0^2 ) / ( v_0 k_p )
\nonumber \\
&&
+ k_{eg}^2 \textbf{E} \cdot \textbf{P} / v_0 + \textbf{g} \cdot \textbf{p} / v_0 ~,
\end{eqnarray}
where $W_{10}^i \approx k_p p_0 v_0$, $\textbf{W}_1 \approx k_p \textbf{p} v_0$, $\textbf{W}_2 \approx k_p \textbf{P} v_0$. $\textbf{f}^* \circ \textbf{f} = - ( \textbf{b} \cdot \textbf{b} - \textbf{g} \cdot \textbf{g} / v_0^2 )$ is the quaternion norm of the gravitational strength $\textbf{f}$. Meanwhile $\textbf{F}^* \circ \textbf{F} = - ( \textbf{B} \cdot \textbf{B} - \textbf{E} \cdot \textbf{E} / v_0^2 )$ is the $S$-quaternion norm of the electromagnetic strength $\textbf{F}$.

When $N_{10} = 0$, the above will be reduced to the mass continuity equation, including the term capable of translating into the Joule heat, $( - k_{eg}^2 \textbf{E} \cdot \textbf{P} - \textbf{g} \cdot \textbf{p} )$. It means that the field is able to release the heat to the particle, or absorb the heat from the particle. The heat exchange between the field and the particle is associated with the field strength and particle velocity, impacting the mass continuity equation directly.

Under the extreme conditions there is no field strength, the above can be simplified into the mass continuity equation in the classical field theory,
\begin{equation}
\partial_0 p_0 - \nabla \cdot \textbf{p} = 0 ~.
\end{equation}

Similarly in case the field strength is comparative weak, the force density,
\begin{eqnarray}
\textbf{N}_1^i = && ( W_{10}^i \textbf{g} / v_0 + \textbf{g} \times \textbf{W}_1^i / v_0
\nonumber \\
&&
- W_{10} \textbf{b} - \textbf{b} \times \textbf{W}_1 ) / v_0
\nonumber \\
&&
+ k_{eg}^2 ( \textbf{E} \circ \textbf{W}_{20}^i / v_0 + \textbf{E} \times \textbf{W}_2^i / v_0
\nonumber \\
&&
- \textbf{B} \circ \textbf{W}_{20} - \textbf{B} \times \textbf{W}_2 ) / v_0
\nonumber \\
&&
- ( \partial_0 \textbf{W}_1 + \nabla W_{10}^i + \nabla \times \textbf{W}_1^i ) ~,
\end{eqnarray}
can be written approximately as,
\begin{eqnarray}
\textbf{N}_1^i / k_p \approx  && - \partial_0 (\textbf{p} v_0)  + p_0 \textbf{g} / v_0
\nonumber \\
&&
+ L_{10} ( \textbf{g} \times \textbf{b} + k_{eg}^2 \textbf{E} \times \textbf{B} ) / ( v_0^2 k_p )
\nonumber \\
&&
- \textbf{b} \times \textbf{p} - \nabla (p_0 v_0)
\nonumber \\
&&
+ k_{eg}^2 ( \textbf{E} \circ \textbf{P}_0 / v_0 - \textbf{B} \times \textbf{P} ) ~,
\end{eqnarray}
where $\textbf{W}_{20}^i \approx k_p \textbf{P}_0 v_0$. $\textbf{W}_2^i \approx \textbf{v} \times \textbf{P} $. $( p_0 \textbf{g} / v_0 )$ is the gravity. $ \partial_0 ( - \textbf{p} v_0)$ is the inertial force. $ \{k_{eg}^2 ( \textbf{E} \circ \textbf{P}_0 / v_0 - \textbf{B} \times \textbf{P} ) \}$ covers the Lorentz force and Coulomb force. $\nabla (p_0 v_0)$ is the energy gradient.

When $\textbf{N}_1^i = 0$, the above will deduce the force equilibrium equation, including the existing force terms in the gravitational and electromagnetic fields, and some new force terms.
$\{ L_{10} (  k_{eg}^2 \textbf{E} \times \textbf{B} ) / ( v_0^2 k_p ) \}$ is in direct proportion to the electromagnetic momentum. Comparing with the force in the classical electromagnetic theory states that $k_{eg}^2 = \mu_g / \mu_e < 0$. The force $\textbf{N}_1^i$ in the above requires that the gravitational field must situate on the quaternion space $\mathbb{H}_g$, meanwhile the electromagnetic field has to stay on the $S$-quaternion space $\mathbb{H}_e$, but not vice versa. Obviously the inertial force term $( - v_0 \partial_0 \textbf{p} )$ plays a crucial role in the space discrimination.

\subsection{\label{sec:level2}Component $\mathbb{N}_e$}

In the $S$-quaternion space $\mathbb{H}_e$, the component $\mathbb{N}_e$ can be written as,
\begin{eqnarray}
\mathbb{N}_e = \emph{i} \textbf{N}_{20}^i + \textbf{N}_{20} + \emph{i} \textbf{N}_2^i + \textbf{N}_2~,
\end{eqnarray}
where $\textbf{N}_{20}$ is able to infer the current continuity equation. $\textbf{N}_2 = \Sigma N_{2k} \emph{\textbf{I}}_k$, $\textbf{N}_{20} = N_{20} \emph{\textbf{I}}_0$. $\textbf{N}_2^i = \Sigma N_{2k}^i \emph{\textbf{I}}_k$, $\textbf{N}_{20}^i = N_{20}^i \emph{\textbf{I}}_0$. $N_{2j}$ and $N_{2j}^i$ are all real numbers. That is,
\begin{eqnarray}
\textbf{N}_{20}^i = && ( \textbf{g} \cdot \textbf{W}_2^i / v_0 - \textbf{b} \cdot \textbf{W}_2 ) / v_0
                  \nonumber \\
                  &&
                  - ( \partial_0 \textbf{W}_{20} + \nabla \cdot \textbf{W}_2^i )
                  \nonumber \\
                  && + ( \textbf{E} \cdot \textbf{W}_1^i / v_0 - \textbf{B} \cdot \textbf{W}_1 ) / v_0 ~,
\\
\textbf{N}_{20} = && ( \textbf{g} \cdot \textbf{W}_2 / v_0 + \textbf{b} \cdot \textbf{W}_2^i ) / v_0
                  \nonumber \\
                  &&
                  + ( \partial_0 \textbf{W}_{20}^i - \nabla \cdot \textbf{W}_2 )
                  \nonumber \\
                  && + ( \textbf{E} \cdot \textbf{W}_1 / v_0 + \textbf{B} \cdot \textbf{W}_1^i ) / v_0 ~,
\\
\textbf{N}_2^i = && ( \textbf{g} \circ \textbf{W}_{20}^i / v_0 + \textbf{g} \times \textbf{W}_2^i / v_0
                  \nonumber \\
                  &&
                  - \textbf{b} \circ \textbf{W}_{20} - \textbf{b} \times \textbf{W}_2 ) / v_0
                  \nonumber \\
                  && + ( W_{10}^i \textbf{E} / v_0 + \textbf{E} \times \textbf{W}_1^i / v_0
                  \nonumber \\
                  &&
                  - W_{10} \textbf{B} - \textbf{B} \times \textbf{W}_1 ) / v_0
                  \nonumber \\
                  && - ( \partial_0 \textbf{W}_2 + \nabla \circ \textbf{W}_{20}^i + \nabla \times \textbf{W}_2^i ) ~,
\\
\textbf{N}_2 = && ( \textbf{g} \circ \textbf{W}_{20} / v_0 + \textbf{g} \times \textbf{W}_2 / v_0
                  \nonumber \\
                  &&
                  + \textbf{b} \circ \textbf{W}_{20}^i + \textbf{b} \times \textbf{W}_2^i ) / v_0
                  \nonumber \\
                  && + ( W_{10} \textbf{E} / v_0 + \textbf{E} \times \textbf{W}_1 / v_0
                  \nonumber \\
                  &&
                  + W_{10}^i \textbf{B} + \textbf{B} \times \textbf{W}_1^i ) / v_0
                  \nonumber \\
                  && + ( \partial_0 \textbf{W}_2^i - \nabla \circ \textbf{W}_{20} - \nabla \times \textbf{W}_2  ) ~.
\end{eqnarray}

In case the field strength is comparative weak, $\textbf{N}_{20}$ is written approximately as,
\begin{eqnarray}
\textbf{N}_{20} / k_p \approx && \partial_0 ( \textbf{P}_0 v_0 ) - \nabla \cdot ( \textbf{P} v_0 )
\nonumber \\
&&
+ \textbf{g} \cdot \textbf{P} / v_0 + \textbf{E} \cdot \textbf{p} / v_0
\nonumber \\
&&
+ ( \textbf{b} \cdot \textbf{b} + \textbf{B} \cdot \textbf{B} ) \textbf{L}_{20} / ( v_0 k_p ) ~,
\end{eqnarray}
where the magnetic flux density $\textbf{B}$ and the gravitational strength component $\textbf{b}$ both make a contribution to the above.

When $\textbf{N}_{20} = 0$, the above is able to reason out the current continuity equation, including the interacting term $ \{ (\textbf{g} \cdot \textbf{P} + \textbf{E} \cdot \textbf{p} ) / v_0 \}$ between the gravitational field and electromagnetic field. It means that the current continuity equation may be disturbed by some influencing factors, such as the field potential.

Under the extreme conditions there is no field strength, the above can be reduced into the current continuity equation in the classical field theory,
\begin{equation}
\partial_0 \textbf{P}_0 - \nabla \cdot \textbf{P} = 0 ~.
\end{equation}

In the classical electromagnetic theory, it requires the current continuity equation to participate the deduction of the Maxwell's equations. However in the field theory described with the octonion, the current continuity equation and the mass continuity equation both are the direct inferences of the field theory. Therefore the current continuity equation really does not need to be taken part in the deduction of the Maxwell's equations anymore. For the charged particle and electric current, it is necessary to satisfy simultaneously the two continuity equations, and then enable those two continuity equations to be correlative each other in this case.

In the octonion space, when $\mathbb{N} = 0$, Eq.(21) can be separated into eight equations, which were not considered to interconnect each other in the past. Any charged particle must obey all of these eight equations simultaneously, when there are the electromagnetic field and the gravitational field.

In the force equilibrium equation, the inertial force, gravity, electromagnetic force, and energy gradient etc make a contribution to the equation. Especially when the energy gradient is much stronger than other force terms, it is able to attract or repulse all kinds of particles of ordinary matter. And the energy gradient caused by the gravitational strength component $\textbf{b}$ may be applied to puzzle about the dynamic of cosmic jets.

In the equation for precessional angular velocity, many torque terms coming from the electromagnetic and gravitational fields, will impact the precessional angular frequency of charged particle, when there are the electromagnetic and gravitational fields simultaneously.

It shows that the definition of octonion force is able to contain more physics contents, which were considered to be independent of each other before, such as the force equilibrium equation, the equation for precessional angular velocity, the mass continuity equation, and the current continuity equation etc (Table IV). The force equilibrium equation is able to conclude the Newton's law of universal gravitation. The equation for precessional angular velocity is capable of deducing the angular velocity of Larmor precession for one charged particle.

\begin{table}[h]
\caption{Some definitions of the physics quantity in the gravitational and electromagnetic fields described with the complex octonion.}
\label{tab:table2}
\centering
\begin{ruledtabular}
\begin{tabular}{lll}
physics~quantity             &   definition                                                                                 \\
\hline
quaternion~operator          &  $\square = i \partial_0 + \Sigma \emph{\textbf{i}}_k \partial_k$                            \\
radius~vector                &  $\mathbb{R} = \mathbb{R}_g + k_{eg} \mathbb{R}_e  $                                         \\
integral~function            &  $\mathbb{X} = \mathbb{X}_g + k_{eg} \mathbb{X}_e  $                                         \\
field~potential              &  $\mathbb{A} = i \square^\times \circ \mathbb{X}  $                                          \\
field~strength               &  $\mathbb{F} = \square \circ \mathbb{A}  $                                                   \\
field~source                 &  $\mu \mathbb{S} = - ( i \mathbb{F} / v_0 + \square )^* \circ \mathbb{F} $                   \\

linear~momentum              &  $\mathbb{P} = \mu \mathbb{S} / \mu_g $                                                      \\
angular~momentum             &  $\mathbb{L} = ( \mathbb{R} + k_{rx} \mathbb{X} )^\times \circ \mathbb{P} $                  \\
octonion~torque              &  $\mathbb{W} = - v_0 ( i \mathbb{F} / v_0 + \square ) \circ \mathbb{L} $                     \\
octonion~force               &  $\mathbb{N} = - ( i \mathbb{F} / v_0 + \square ) \circ \mathbb{W} $                         \\
\end{tabular}
\end{ruledtabular}
\end{table}

\section{\label{sec:level1}Conclusions and Discussions}

The paper introduced the quaternion into the field theory, to describe simultaneously the features of gravitational and electromagnetic fields. The space extended from the electromagnetic field is independent of that from the gravitational field. Meanwhile the spaces of gravitational field and of electromagnetic field can be chosen as the quaternion spaces. Furthermore the components of those two spaces and of relevant physics quantities may be complex numbers.

In the gravitational and electromagnetic field theories, the scalar parts of quaternion operator and of field potential are all imaginary numbers. However this simple case is still able to result in many components of other physics quantities to become the imaginary numbers and even complex numbers. In order to achieve the field theory described with the quaternion, which approximates to the classical field theory, it is necessary to choose suitable gauge conditions. Therefore it is able to be said that the gauge condition is also the essential ingredient for the field theory.

In the quaternion space, it is capable of deducing the angular momentum, torque, force, energy, and mass continuity equation etc. The mass continuity equation is one indispensable part of the gravitational theory described with the quaternion. Under the condition of weak gravitational strength, the mass continuity equation described with the quaternion can be degenerated into that in the classical gravitational theory. The force includes the inertial force, gravity, electromagnetic force, and energy gradient etc in the classical field theory.

According to the multiplication of octonion, among the product of two physics quantities in the $S$-quaternion space, some terms may stay on the $S$-quaternion space still, while other terms will situate on the quaternion space, such as the electromagnetic force. In the $S$-quaternion space, it is able to reason out directly the dipole moment and current continuity equation etc. The current continuity equation is one intrinsic part of the electromagnetic theory described with the $S$-quaternion. In the case of weak electromagnetic strength, the current continuity equation described with the $S$-quaternion can be simplified into that in the classical electromagnetic theory.

It should be noted that the paper discussed only some simple cases about the torque, force, energy, force equilibrium equation, and continuity equations etc in the field theory described with the quaternion/$S$-quaternion. However it clearly states that the complex octonion is able to availably describe the physics features of electromagnetic and gravitational fields. This will afford the theoretical basis for further analysis, and is helpful to research the property of the force and continuity equations in the following researches.

\begin{table*}
\caption{The analogy between the physics quantity of gravitational field with that of electromagnetic field in the complex octonion space.}
\begin{ruledtabular}
\begin{tabular}{lll}

physics~quantity             &  gravitational~field                                                                &  electromagnetic~field                                       \\
\hline
field~potential              &  $a_0$, gravitational~scalar~potential                                              &  $\textbf{A}_0$, electromagnetic~scalar~potential            \\
                             &  $\textbf{a}$, (similar to $\textbf{A}$)                                            &  $\textbf{A}$, electromagnetic~vector~potential              \\
field~strength	             &  $\textbf{g}$, gravitational~acceleration 	                                       &  $\textbf{E}$, electric~field~intensity                      \\
                  	         &  $\textbf{b}$, (similar to $\textbf{B}$)                                            &  $\textbf{B}$, magnetic~flux~density                         \\
field~source	             &  $s_0$, mass~density 	                                                           &  $\textbf{S}_0$, electric~charge~density                     \\
         	                 &  $\textbf{s}$, linear~momentum~density 	                                           &  $\textbf{S}$, electric~current~density                      \\

angular~momentum	         &  $L_{10}$, dot~product                                                              &  $\textbf{L}_{20}$, (similar to $L_{10}$)                    \\
	                         &  $\textbf{L}_1$, angular~momentum 	                                               &  $\textbf{L}_2$, magnetic~dipole~moment                      \\
	                         &  $\textbf{L}_1^i$, (similar to $\textbf{L}_2^i$)                                    &  $\textbf{L}_2^i$, electric~dipole~moment                    \\
octonion~torque	             &  $W_{10}$, divergence~of 	                                                       &  $\textbf{W}_{20}$, divergence~of                            \\
                             &  ~~~~~~~angular~momentum                                                            &  ~~~~~~~~magnetic~dipole~moment                              \\
	                         &  $W_{10}^i$,	energy                                                                 &  $\textbf{W}_{20}^i$, (similar to $W_{10}^i$)                \\
	                         &  $\textbf{W}_1$,	curl~of~angular~momentum                                           &  $\textbf{W}_2$, curl~of~magnetic~dipole                     \\
                             &                                                                                     &  ~~~~~~~~moment,~and~derivative                              \\
                             &                                                                                     &  ~~~~~~~~of~electric~dipole~moment                           \\
	                         &  $\textbf{W}_1^i$, torque                                                           &  $\textbf{W}_2^i$, (similar to $\textbf{W}_1^i$)             \\
octonion~force     	         &  $N_{10}$, power 	                                                               &  $\textbf{N}_{20}$, (similar to $N_{10}$)                    \\
	                         &  $N_{10}^i$, torque~divergence                                                      &  $\textbf{N}_{20}^i$, (similar to $N_{10}^i$)                \\
	                         &  $\textbf{N}_1$,	torque~derivative                                                  &  $\textbf{N}_2$, (similar to $\textbf{N}_1$)                 \\
	                         &  $\textbf{N}_1^i$, force                                                            &  $\textbf{N}_2^i$, (similar to $\textbf{N}_1^i$)             \\

\end{tabular}
\end{ruledtabular}
\end{table*}

\begin{acknowledgements}
The author is indebted to the anonymous referee for their valuable and constructive comments on the previous manuscript. This project was supported partially by the National Natural Science Foundation of China under grant number 60677039.
\end{acknowledgements}

\appendix

\section{Maxwell's equations}

The definition of octonion field source in Eq.(3) can be expressed as,
\begin{eqnarray}
&&- \square^* \circ ( \mathbb{F}_g + k_{eg} \mathbb{F}_e )  - ( \emph{i} \mathbb{F} / v_0 )^* \circ \mathbb{F}
\nonumber \\
= && \mu_g \mathbb{S}_g + k_{eg} \mu_e \mathbb{S}_e - ( \emph{i} \mathbb{F} / v_0 )^* \circ \mathbb{F} ~,
\end{eqnarray}
further the above can be separated into two equations, Eqs.(3) and (4), according to the coefficient $k_{eg}$ and the basis vectors, $\mathbb{H}_g$ and $\mathbb{H}_e$ .

In the quaternion space $\mathbb{H}_g$, the definition of gravitational source in Eq.(2) can be expressed as,
\begin{eqnarray}
- \mu_g ( \emph{i} s_0 + \textbf{s} ) = ( \emph{i} \partial_0 + \nabla )^* \circ ( \emph{i} \textbf{g} / v_0 + \textbf{b} )  ~.
\end{eqnarray}

Comparing both sides of the equal sign in the above will yield,
\begin{eqnarray}
&& \nabla^* \cdot \textbf{b} = 0 ~,
\\
&& \partial_0 \textbf{b} + \nabla^* \times \textbf{g} / v_0 = 0 ~,
\\
&& \nabla^* \cdot \textbf{g} / v_0 = - \mu_g s_0   ~,
\\
&& \nabla^* \times \textbf{b} - \partial_0 \textbf{g} / v_0 = - \mu_g \textbf{s} ~.
\end{eqnarray}

Eqs.(A3)-(A6) are the gravitational field equations. Because the gravitational constant $G$ is weak and the velocity ratio $\textbf{v} / c$ is tiny, the gravity and $\textbf{b}$ produced by the linear momentum $\textbf{s}$ can be ignored in general. When $\textbf{b} = 0$ and $\textbf{a} = 0$, Eq.(A5) will be degenerated into the Poisson equation for the Newton's law of universal gravitation in the classical gravitational theory.

The deduction approach of the gravitational field equations can be used as a reference to be extended to that of electromagnetic field equations.

In the $S$-quaternion space $\mathbb{H}_e$, the definition of electromagnetic source in Eq.(4) can be written as,
\begin{eqnarray}
- \mu_e ( \emph{i} \textbf{S}_0 + \textbf{S} ) = ( \emph{i} \partial_0 + \nabla )^* \circ ( \emph{i} \textbf{E} / v_0 + \textbf{B} )  ~.
\end{eqnarray}

In the above, comparing both sides of the equal sign will reason out the electromagnetic field equations,
\begin{eqnarray}
&& \nabla^* \cdot \textbf{B} = 0 ~,
\\
&& \partial_0 \textbf{B} + \nabla^* \times \textbf{E} / v_0 = 0 ~,
\\
&& \nabla^* \cdot \textbf{E} / v_0 = - \mu_g \textbf{S}_0   ~,
\\
&& \nabla^* \times \textbf{B} - \partial_0 \textbf{E} / v_0 = - \mu_g \textbf{S} ~.
\end{eqnarray}

Comparing Eqs.(A8)-(A11) with the Maxwell's equations in the classical electromagnetic theory states that both of them are equivalent to each other, including the direction of displacement current. It means that in the electromagnetic theory described with the $S$-quaternion, it is still able to reason out the Maxwell's equations in the classical electromagnetic theory, without the participation of the current continuity equation.

From the definition of field strength Eq.(1) and the field equations Eq.(2), it is able to obtain the d'Alembert equation for the gravitational and electromagnetic fields. The equation claims that the octonion field potential $\mathbb{A}$ is determined by the octonion field source $\mathbb{S}$ . That is, the gravitational potential, $ a_0 $ and $ \textbf{a} $, are dependent on the mass and linear momentum, while the electromagnetic potential, $ \textbf{A}_0 $ and $ \textbf{A} $, are dependent on the electric charge and current.

The meaning of field potential in the electromagnetic field in the paper remains the same as that in the classical electromagnetic theory. In the octonion space, the gravitational potential possesses the scalar part and vector part. When the velocity ratio $(\textbf{v} / c)$ in the quaternion space, $\mathbb{H}_g$, is quite tiny, the gravitational potential in the paper will be reduced into that in the classical gravitational theory, in which the vector part of gravitational potential is equal to zero.

\section{Dipole moment}

The octonion angular momentum density in Eq.(8) can be rewritten as,
\begin{equation}
\mathbb{L} = \mathbb{L}_g + k_{eg} \mathbb{L}_e  ~,
\end{equation}
further the above can be separated into two parts, Eqs.(9) and (10), according to the basis vectors, $\mathbb{H}_g$ and $\mathbb{H}_e$ .

In the quaternion space $\mathbb{H}_g$, the part $\mathbb{L}_g$ in Eq.(9) can be expressed as,
\begin{equation}
\mathbb{L}_g = L_{10} + \emph{i} \textbf{L}_1^i + \textbf{L}_1  ~,
\end{equation}
where $L_{10}$ is relevant to the magnitude of spin angular momentum. $\textbf{L}_1^i$ is similar to the electric dipole moment. $\textbf{L}_1$ includes the orbital angular momentum density.

Further the above components are expanded into,
\begin{eqnarray}
L_{10} = &&  (r_0 + k_{rx} x_0) p_0 + (\textbf{r} + k_{rx} \textbf{x}) \cdot \textbf{p}
\nonumber
\\
&& + k_{eg}^2 \{ (\textbf{R}_0 + k_{rx} \textbf{X}_0) \circ \textbf{P}_0
\nonumber
\\
&&
+ (\textbf{R} + k_{rx} \textbf{X})  \cdot \textbf{P} \}
~,
\\
\textbf{L}_1^i = &&  p_0 (\textbf{r} + k_{rx} \textbf{x}) - (r_0 + k_{rx} x_0) \textbf{p}
\nonumber
\\
&& +  k_{eg}^2 \{ (\textbf{R} + k_{rx} \textbf{X}) \circ \textbf{P}_0
\nonumber
\\
&&
- (\textbf{R}_0 + k_{rx} \textbf{X}_0) \circ \textbf{P} \}
~,
\\
\textbf{L}_1 =  && (\textbf{r} + k_{rx} \textbf{x}) \times \textbf{p}
\nonumber
\\
&&
+ k_{eg}^2 (\textbf{R} + k_{rx} \textbf{X})  \times \textbf{P} ~,
\end{eqnarray}
where $( \textbf{r} \times \textbf{p} )$ is the orbital angular momentum in the classical field theory. And $( - \textbf{r} \cdot \textbf{p} )$ may be the magnitude of spin angular momentum.

In the $S$-quaternion space $\mathbb{H}_e$, the part $\mathbb{L}_e$ in Eq.(10) can be written as,
\begin{eqnarray}
\mathbb{L}_e = \textbf{L}_{20} + \emph{i} \textbf{L}_2^i + \textbf{L}_2  ~,
\end{eqnarray}
where $\textbf{L}_{20}$ is relevant to the spin magnetic moment. $\textbf{L}_2^i$ covers the electric dipole moment etc. $\textbf{L}_2$ includes the magnetic dipole moment etc.

Further the above parts are expanded into,
\begin{eqnarray}
\textbf{L}_{20} = &&  (r_0 + k_{rx} x_0) \textbf{P}_0 + (\textbf{r} + k_{rx} \textbf{x}) \cdot \textbf{P}
\nonumber
\\
&&
+  p_0 (\textbf{R}_0 + k_{rx} \textbf{X}_0)
+ (\textbf{R} + k_{rx} \textbf{X}) \cdot \textbf{p} ~,
\\
\textbf{L}_2^i = &&  - (r_0 + k_{rx} x_0) \textbf{P} + (\textbf{r} + k_{rx} \textbf{x}) \circ \textbf{P}_0
\nonumber
\\
&&
- (\textbf{R}_0 + k_{rx} \textbf{X}_0) \circ \textbf{p}
+ p_0 (\textbf{R} + k_{rx} \textbf{X}) ~,
\\
\textbf{L}_2 = &&  (\textbf{r} + k_{rx} \textbf{x}) \times \textbf{P}
+ (\textbf{R} + k_{rx} \textbf{X}) \times \textbf{p} ~,
\end{eqnarray}
where $( - k_{eg} \textbf{r} \cdot \textbf{P} )$ may be the spin magnetic moment, with the basis vector $\emph{\textbf{I}}_0$ . $( - k_{eg} \textbf{r} \times \textbf{P} )$ is the magnetic dipole moment, while $( k_{eg} \textbf{r} \circ \textbf{P}_0 )$ is the electric dipole moment.

\section{Torque}

The definition of octonion torque density in Eq.(11) can be rewritten as,
\begin{eqnarray}
\mathbb{W} = && - v_0 \{ \emph{i} ( \mathbb{F}_g + k_{eg} \mathbb{F}_e ) / v_0 + \square \}
\nonumber
\\
&&
\circ ( \mathbb{L}_g + k_{eg} \mathbb{L}_e ) ~,
\end{eqnarray}
further the octonion torque density in Eq.(12) will be separated into two parts, $\mathbb{W}_g$ and $\mathbb{W}_e$, according to the basis vectors, $\mathbb{H}_g$ and $\mathbb{H}_e$ . That is,
\begin{equation}
\mathbb{W} = \mathbb{W}_g + k_{eg} \mathbb{W}_e ~.
\end{equation}

In the quaternion space $\mathbb{H}_g$, the part $\mathbb{W}_g$ in Eq.(13) can be expressed as,
\begin{equation}
\mathbb{W}_g = \emph{i} W_{10}^i + W_{10} + \emph{i} \textbf{W}_1^i + \textbf{W}_1 ~,
\end{equation}
where $W_{10}^i$ is the energy density. $W_{10}$ is the divergence of angular momentum density. $(-\textbf{W}_1^i)$ is the torque density. $\textbf{W}_1$ is the curl of angular momentum density.

Further the above parts are expanded into,
\begin{eqnarray}
W_{10}^i = &&  ( \textbf{g} \cdot \textbf{L}_1^i / v_0 - \textbf{b} \cdot \textbf{L}_1 )
\nonumber
\\
&&
- v_0 (  \partial_0 L_{10} +  \nabla \cdot \textbf{L}_1^i)
\nonumber
\\
&&
+ k_{eg}^2 ( \textbf{E} \cdot \textbf{L}_2^i / v_0 - \textbf{B} \cdot \textbf{L}_2 )
~,
\\
W_{10} =
&&
( \textbf{b} \cdot \textbf{L}_1^i + \textbf{g} \cdot \textbf{L}_1 / v_0 ) - v_0 \nabla \cdot \textbf{L}_1
\nonumber
\\
&&
+ k_{eg}^2 ( \textbf{B} \cdot \textbf{L}_2^i + \textbf{E} \cdot \textbf{L}_2 / v_0 ) ~,
\\
\textbf{W}_1^i =
&&
( \textbf{g} \times \textbf{L}_1^i / v_0 - L_{10} \textbf{b} - \textbf{b} \times \textbf{L}_1 )
\nonumber
\\
&&
- v_0 (  \partial_0 \textbf{L}_1 +  \nabla \times \textbf{L}_1^i )
\nonumber \\
&& + k_{eg}^2 ( \textbf{E} \times \textbf{L}_2^i / v_0 - \textbf{B} \circ \textbf{L}_{20}
\nonumber \\
&&
- \textbf{B} \times \textbf{L}_2)
~,
\\
\textbf{W}_1 =
&&
( \textbf{g} L_{10} / v_0 + \textbf{g} \times \textbf{L}_1 / v_0 + \textbf{b} \times \textbf{L}_1^i )
\nonumber
\\
&&
- v_0 ( - \partial_0 \textbf{L}_1^i + \nabla L_{10} +  \nabla \times \textbf{L}_1 )
\nonumber
\\
&& + k_{eg}^2 ( \textbf{E} \circ \textbf{L}_{20} / v_0 + \textbf{E} \times \textbf{L}_2 / v_0
\nonumber
\\
&&
+ \textbf{B} \times \textbf{L}_2^i )
~,
\end{eqnarray}
where $\{- k_{eg}^2 (\textbf{E} / v_0 ) \times ( \textbf{r} \circ \textbf{P}_0 ) \}$ is the torque term between the electric field intensity with the electric dipole moment. $\{ k_{eg}^2 \textbf{B} \times ( \textbf{r} \times \textbf{P} ) \}$ is the torque term between the magnetic flux density with the magnetic dipole moment. $\{ k_{eg}^2 \textbf{B} \circ ( \textbf{r} \cdot \textbf{P} ) \}$ may be the torque term between the magnetic flux density with the spin magnetic moment.

In the $S$-quaternion space $\mathbb{H}_e$, the part $\mathbb{W}_e$ can be written as Eq.(16), which is expanded into Eqs.(17)-(20).

\section{Force}

The definition of octonion torque density in Eq.(21) can be rewritten as,
\begin{eqnarray}
\mathbb{N} = && - \{ \emph{i} ( \mathbb{F}_g + k_{eg} \mathbb{F}_e ) / v_0 + \square \}
\nonumber \\
&&
\circ ( \mathbb{W}_g + k_{eg} \mathbb{W}_e ) ~,
\end{eqnarray}
further the octonion torque density in Eq.(22) will be separated into two parts, $\mathbb{N}_g$ and $\mathbb{N}_e$, according to the basis vectors, $\mathbb{H}_g$ and $\mathbb{H}_e$ . That is,
\begin{equation}
\mathbb{N} = \mathbb{N}_g + k_{eg} \mathbb{N}_e ~.
\end{equation}

In the quaternion space $\mathbb{H}_g$, the part $\mathbb{N}_g$ in Eq.(23) can be expressed as,
\begin{equation}
\mathbb{N}_g = \emph{i} N_{10}^i + N_{10} + \emph{i} \textbf{N}_1^i + \textbf{N}_1 ~,
\end{equation}
where $N_{10}^i$ is the torque divergence. $N_{10}$ is the power density. $\textbf{N}_1^i$ is the force density. $\textbf{N}_1$ is the torque derivative.

Further the above parts are expanded into,
\begin{eqnarray}
N_{10}^i = &&  ( \textbf{g} \cdot \textbf{W}_1^i / v_0 - \textbf{b} \cdot \textbf{W}_1 ) / v_0
\nonumber \\
&&
- ( \partial_0 W_{10} + \nabla \cdot \textbf{W}_1^i )
\nonumber \\
&&
+ k_{eg}^2 ( \textbf{E} \cdot \textbf{W}_2^i / v_0 - \textbf{B} \cdot \textbf{W}_2 ) / v_0  ~,
\\
N_{10} = &&  ( \textbf{g} \cdot \textbf{W}_1 / v_0 + \textbf{b} \cdot \textbf{W}_1^i ) / v_0
\nonumber \\
&&
+ ( \partial_0 W_{10}^i -  \nabla \cdot \textbf{W}_1 )
\nonumber \\
&&
+ k_{eg}^2 ( \textbf{E} \cdot \textbf{W}_2/ v_0 + \textbf{B} \cdot \textbf{W}_2^i ) / v_0  ~  ,
\\
\textbf{N}_1^i = && ( W_{10}^i \textbf{g} / v_0 + \textbf{g} \times \textbf{W}_1^i / v_0
\nonumber \\
&&
- W_{10} \textbf{b} - \textbf{b} \times \textbf{W}_1 ) / v_0
\nonumber \\
&& + k_{eg}^2 ( \textbf{E} \circ \textbf{W}_{20}^i / v_0 + \textbf{E} \times \textbf{W}_2^i / v_0
\nonumber \\
&&
- \textbf{B} \circ \textbf{W}_{20} - \textbf{B} \times \textbf{W}_2 ) / v_0
\nonumber \\
&& - ( \partial_0 \textbf{W}_1 + \nabla W_{10}^i + \nabla \times \textbf{W}_1^i )  ~,
\\
\textbf{N}_1 = &&  ( W_{10} \textbf{g} / v_0 + \textbf{g} \times \textbf{W}_1 / v_0
\nonumber \\
&&
+ W_{10}^i \textbf{b} + \textbf{b} \times \textbf{W}_1^i ) / v_0
\nonumber \\
&& + k_{eg}^2 ( \textbf{E} \circ \textbf{W}_{20} / v_0 + \textbf{E} \times \textbf{W}_2/ v_0
\nonumber \\
&&
+ \textbf{B} \circ \textbf{W}_{20}^i + \textbf{B} \times \textbf{W}_2^i ) / v_0
\nonumber \\
&& + ( \partial_0 \textbf{W}_1^i - \nabla W_{10} - \nabla \times \textbf{W}_1 )  ~,
\end{eqnarray}
where the energy gradient, $(- \nabla W_{10}^i)$, is able to exert the force on the ordinary matter, including the neutral or charged particles etc.

In case the field strength is comparative weak, the term, $\textbf{N}_1$, can be degenerated approximately to,
\begin{eqnarray}
\textbf{N}_1 / k_p \approx &&  - v_0 \nabla \times \textbf{p}  + p_0 \textbf{b}   + \textbf{g} \times \textbf{p} / v_0
\nonumber \\
&& + k_{eg}^2 \{ \textbf{E} \times \textbf{P} / v_0  + \textbf{B} \circ \textbf{P}_0
\nonumber \\
&&
+ \textbf{B} \times (\textbf{v} \times \textbf{P} ) / ( k_p v_0 ) \} ~.
\end{eqnarray}

When $\textbf{N}_1 = 0$ and there is only the gravitational strength component $\textbf{b}$, the above will be reduced to
\begin{eqnarray}
- v_0 \nabla \times \textbf{p} + p_0 \textbf{b}  = 0  ~.
\end{eqnarray}

The linear momentum is, $\textbf{p} = m \textbf{v}$ , for one neutral particle. The velocity curl is, $ \nabla \times \textbf{v} = 2 \overrightarrow{\omega}_1 $ , when $k = 2$. And $\overrightarrow{\omega}_1$ is the precessional angular velocity. The above will yield,
\begin{eqnarray}
\textbf{b} =  2 \overrightarrow{\omega}_1  ~.
\end{eqnarray}

It means that the gravitational strength component $\textbf{b}$ corresponds to the double of the precessional angular velocity in the gravitational field when $k = 2$ . Meanwhile the gravitational acceleration $\textbf{g}$ corresponds to the linear acceleration from Eq.(28).

When $\textbf{N}_1 = 0$ and there is only the magnetic flux density $\textbf{B}$, Eq.(D8) will be reduced to
\begin{eqnarray}
 - v_0 \nabla \times \textbf{p} + k_{eg}^2 \textbf{B} \circ \textbf{P}_0  = 0  ~.
\end{eqnarray}

For one charged particle, there are, $k_{eg}^2 \textbf{P}_0 = q V_0 \emph{\textbf{I}}_0$ , and $\textbf{p} = m \textbf{v}$ . The velocity curl is, $ \nabla \times \textbf{v} = 2 \overrightarrow{\omega}_2 $ , when $k = 2$ . In the magnetic field $\textbf{B}$, the above is able to infer the angular frequency of Larmor precession for the charged particle,
\begin{eqnarray}
\omega_2 = - B q / ( 2 m ) ~,
\end{eqnarray}
where $\omega_2$ and $B$ are the magnitude of $\overrightarrow{\omega}_2$ and $\textbf{B}$ respectively. $(V_0 / v_0)$ is equal to 1 in general.

According to Eq.(D8), not only the gravitational acceleration $\textbf{g}$ but also the electric field intensity $\textbf{E}$ has an influence on the precessional angular velocity.

In the $S$-quaternion space $\mathbb{H}_e$, the part $\mathbb{N}_e$ can be written as Eq.(29), which is expanded into Eqs.(30)-(33).

{}

\end{document}